\documentclass[twocolumn,prl,aps,nobalancelastpage,amssymb]{revtex4}

\usepackage{graphicx}

\begin{document}
\title{Directional field-induced metallization of quasi-one-dimensional Li$_{0.9}$Mo$_6$O$_{17}$}

\author{X.~Xu$^1$, A.~F.~Bangura$^1$, J.~G.~Analytis$^{1,2}$, J.~D.~Fletcher$^1$, M.~M.~J.~French$^1$, N.~Shannon$^1$, J. He$^3$, S. Zhang$^3$, D. Mandrus$^4$, R.~Jin$^{4,5}$ and N.~E.~Hussey$^1$}
\affiliation{$^1$H. H. Wills Physics Laboratory, University of Bristol, Tyndall Avenue, BS8 1TL, U.K.}
\affiliation{$^2$Geballe Laboratory for Advanced Materials, Stanford University, Stanford, California 94305-4045, USA.}
\affiliation{$^3$Department of Physics and Astronomy, Clemson University, Clemson, SC 29631 USA}
\affiliation{$^4$Materials Science and Technology Division, Oak Ridge National Laboratory, Oak Ridge, TN 37831}
\affiliation{$^5$Department of Physics and Astronomy, University of Tennessee, Knoxville, TN 37996}
\date{\today}

\begin{abstract}
We report a detailed magnetotransport study of the highly anisotropic quasi-one-dimensional oxide Li$_{0.9}$Mo$_6$O$_{17}$ whose in-chain electrical resistivity diverges below a temperature $T_{\rm min} \sim$ 25 K. For $T < T_{\rm min}$, a magnetic field applied parallel to the conducting chain induces a large negative magnetoresistance and ultimately, the recovery of a metallic state. We show evidence that this insulator/metal crossover is a consequence of field-induced suppression of a density-wave gap in a highly one-dimensional conductor. At the highest fields studied, there is evidence for the possible emergence of a novel superconducting state with an onset temperature \mbox{$T_c >$ 10 K}.
\end{abstract}

\maketitle

Low-dimensional interacting electron systems provide a rich playground for physicists due to the wide variety of quantum ground states that they exhibit, particularly when these states can be manipulated and explored under accessible laboratory conditions. Magnetic fields, for example, have proved an invaluable tool in the study of quasi-one-dimensional (quasi-1D) systems in which the Zeeman or orbital effects are comparable with the interchain coupling strength $t_\perp$. Indeed, magnetotransport studies on both organic and inorganic quasi-1D compounds have uncovered a wealth of novel phenomena including field-induced spin density wave (SDW) cascades \cite{JeromeSchulz}, dimensional crossovers \cite{Hussey02}, charge density wave (CDW) suppression \cite{Graf04} and CDW enhancement \cite{Balseiro}.

Here we report the observation of a new field-induced phenomenon in a quasi-1D system, namely a highly orientation-dependent insulator/metal crossover in the molybdenum oxide bronze Li$_{0.9}$Mo$_6$O$_{17}$ (LMO). LMO is metallic at room temperature, semiconducting below a temperature $T_{\rm min} \sim$ 25 K and superconducting below $T_c \sim$ 1.8 K \cite{Greenblatt84}. The origin of the resistive upturn below $T_{\rm min}$ has not yet been resolved, with CDW formation, SDW formation and strong localization all put forward as possible origins of the metal-to-insulator crossover. With a magnetic field applied parallel to the most conducting direction ({\bf H}$\| b$), a large negative magnetoresistance (MR) is observed which ultimately restores the metallic state. Analysis shows that this novel insulator/metal crossover is driven by Zeeman splitting of the gap associated with an ordered state, presumably a CDW or SDW. Whilst Zeeman splitting is essentially isotropic, any field component transverse to the chains is found to have a secondary, competing effect on the density wave gap that acts to preserve the insulating state. At the highest fields ({\bf H}$\| b$), the resistivity drops sharply, heralding a possible transition into a novel, highly unidirectional superconducting state with an onset temperature that is significantly enhanced compared with the zero-field value.

\begin{figure*}
\includegraphics[width=6.5cm,angle=270,keepaspectratio=true]{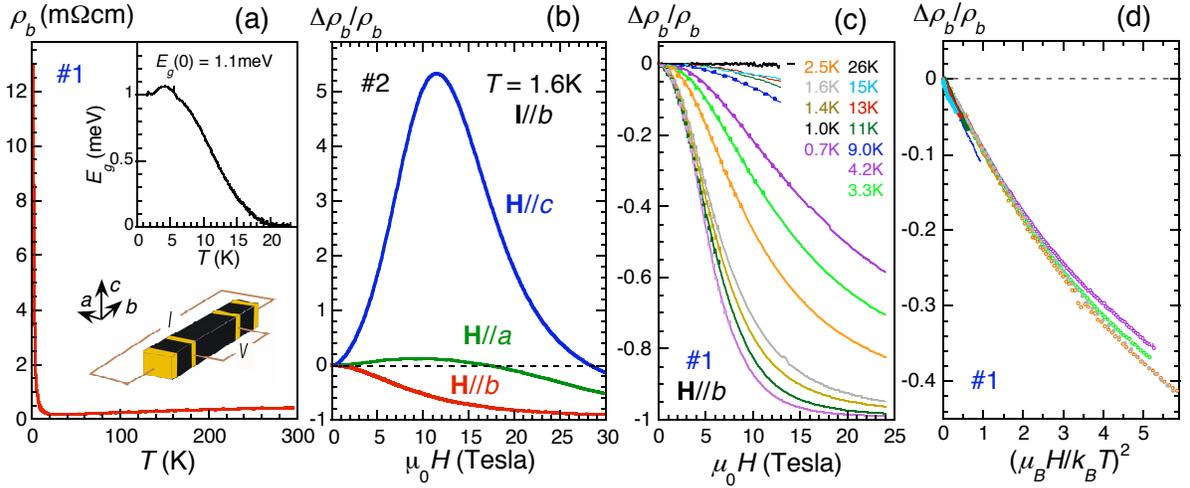}
\caption{(a) $T$-dependence of the in-chain resistivity $\rho_b(T)$ of Li$_{0.9}$Mo$_6$O$_{17}$. Inset: $T$-dependence of the gap
magnitude $E_g$ extracted from fitting the low-$T$ data to $\rho_b$($T$) = $\rho_0$ exp($E_g(T)$/2$k_BT$). Also shown is a
schematic of the electrical contacts. (b) Magnetic field dependence of $\rho_b(T)$ of sample $\#$2 for the three orthogonal field orientations at $T$ = 1.6 K. (c) Solid lines: $\Delta \rho_{b}/\rho_{b}$ for sample $\#$1 with {\bf H}$\| b$ at different temperatures as indicated. Dots: Representative fits of the low-field ($\mu_0H < 15$ T) MR curves to Eq. 1. (d) $\Delta \rho_{b}/\rho_{b}$ versus $(\mu_B H/k_B T)^2$ showing the excellent scaling of the data between 1.6 K and 15 K.} \label{Fig1}
\end{figure*}

LMO crystals were grown by a temperature gradient flux technique \cite{McCarroll84} and cut into bar-shaped samples with the longest dimension parallel to the $b$-axis. Gold contacts were then sputtered onto the crystals in a configuration designed to minimize any possible voltage drop across the least-conducting axes, as shown schematically in Fig.~\ref{Fig1}a. The MR measurements were performed for different field orientations up to 13 T in Bristol, up to 30 T at the HMFL in Nijmegen, Holland and up to 45 T at the NHMFL in Tallahassee.

Fig.~1a shows the in-chain electrical resistivity $\rho_b(T)$ of one of the LMO crystals used in this MR study, labelled hereafter $\#$1. Whilst the $T$-dependence is typical of those reported in the literature, the absolute value of $\rho_b$ at room temperature (0.4 m$\Omega$cm) is significantly lower than previous reports. We attribute this to the improved shorting out of the orthogonal current paths. The corresponding dc electrical anisotropy \mbox{($\rho_c > \rho_a \sim 100\rho_b$)} is comparable with the optical anisotropy reported by Choi {\it
et al.}~\cite{Choi04}, thus confirming the extreme one-dimensionality of LMO. For \mbox{$T_{\rm min} \leq T \leq 300$}~K, $\rho_b(T)$ is approximately \mbox{$T$-linear}.  Below $T_{\rm min}$, $\rho_b(T)$ follows activated-like behavior associated with a $T$-dependent gap of order 1 meV at 0 K, as shown in the inset to Fig.~\ref{Fig1}a. We note that this gap value lies below the far-infrared frequency cut-off of optical measurements reported thus far on LMO \cite{Choi04, Degiorgi} and is therefore consistent with the absence of any gap-like features in the optical response below $T_{\rm min}$.

No trace of superconductivity could be detected in zero field in either of the samples featured in this report down to 0.6~K. In earlier work, Matsuda {\it et al.} showed that $T_c$ in LMO correlated with the magnitude of the resistivity upturn, as defined by the ratio $\rho$(2K)/$\rho$($T_{\rm min}$) \cite{Matsuda}. According to their measurements, values of the ratio larger than 10 are sufficient to halve the zero-field $T_c$. For our crystals, $\rho$(2K)/$\rho$($T_{\rm min}$) $>$ 20 and $T_c <$ 0.6 K, consistent with this correlation, the origin of which is not yet clear. On one hand, it implies some sort of disorder-induced suppression of $T_c$, e.g due to slight off-stoichiometry. On the other, it could also be related to enhanced density wave formation; the larger the upturn, the more Fermi surface gapping there is and hence the less density of states available for superconductivity. A similar correlation has been reported in (TMTSF)$_2$PF$_6$ \cite{Lee02} and attributed to the competition between SDW and superconductivity. This interpretation in terms of competing states is also consistent with the recovery of metallicity (and perhaps superconductivity) in LMO in an applied field, to be described in more detail below.

For $T > T_{\rm min}$, the MR response in LMO is typical of a metal; the orbital in-chain MR (current {\bf I}$\perp${\bf H}) is small, positive and varies quadratically with the applied field $H$, whilst the longitudinal MR ({\bf I}$\|${\bf H}$\| b$) is negligibly small. Below $T_{\rm min}$ however, the MR response undergoes a dramatic transformation and becomes strongly temperature dependent.  The low-$T$ behavior is encapsulated in Fig.~1b where the MR response (plotted as $\Delta \rho_{b}/\rho_{b}$) of a second crystal ($\#$2) at $T$ = 1.6 K is reproduced.

For {\bf H}$\| b$, $\Delta \rho_{b}/\rho_{b}$ is large and negative, with $\rho_{b}$ falling by one order of magnitude in 30 T. (Similar data were observed on three other crystals, including one which superconducted at zero-field). Significantly, $\Delta \rho(H)$ is {\it not} exponential in field (see Fig.~1c) as would be expected were the resistivity upturn below $T_{\rm min}$ due to strong localization~\cite{Gershenson97}. For {\bf H}$\| c$, the low-field MR is large and positive; $\rho_b$ increasing five-fold for $\mu_0H$ = 10~T. The MR curve then goes through a sharp maximum and at the highest fields measured, becomes negative. The MR response for {\bf H}$\| a$ is intermediate between these two extremes. Note that at the highest fields, $\Delta \rho_{b}/\rho_{b}$ is negative for {\it all} orientations of the magnetic field.

Qualitatively similar, though less spectacular, anisotropic MR behavior was recently reported in the quasi-1D cuprate PrBa$_2$Cu$_4$O$_8$ (Pr124) and attributed to the formation of quasi-1D metallic islands sandwiched between strong back-scattering impurities \cite{Narduzzo07}. In Pr124 however, the upturns in $\rho_b(T)$ are relatively modest and compatible with localization \cite{Enayati-Rad07}. Moreover, in contrast to LMO, the MR response in Pr124 does not vary markedly across $T_{\rm min}$ \cite{Narduzzo07}. We therefore believe that the origin of the MR in both systems is distinct. Indeed, the highly non-monotonic field dependence in LMO for {\bf I}$\perp${\bf H} suggests there are two competing mechanisms at work here, one that drives the system more insulating and one that destroys the insulating behavior.

\begin{figure*}
\includegraphics[width=18cm,keepaspectratio=true]{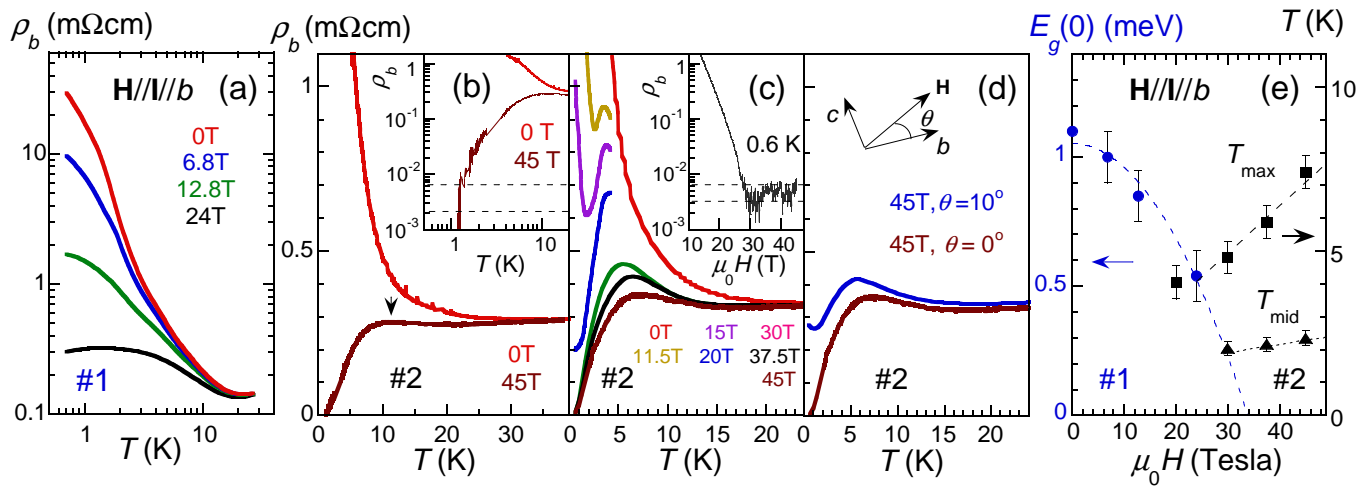}
\caption{{\bf a}) Log-log plot of $\rho_b(T)$ of sample $\#$1 for 0 $\leq \mu_0H \leq$ 24 T. {\bf b}) $\rho_b(T)$ of $\#$2 at $\mu_0 H$ = 0 T (upper red line) and 45 T ({\bf H}$\| b$, lower brown line). The arrow indicates $T_{\rm max}$. Inset: Same data re-plotted on a log-log scale. The horizontal dashed lines indicate the noise floor of our measurements. {\bf c}) $\rho_b(T)$ of $\#$2 for different field values 0 $\leq \mu_0H \leq$ 45 T as indicated \cite{secondrun}. Inset: Field sweep on $\#$2 at $T$ = 0.6 K on a semi-log scale ({\bf H}$\| b$). Again, the horizontal dashed lines indicate the noise floor of our measurements. {\bf d}) $\rho_b(T)$ of $\#$2 for $\theta$ = 0$^{\circ}$ (bottom line) and 10$^{\circ}$ (top line) where $\theta$ is rotation within the $bc$-plane. {\bf e}) Corresponding phase diagram for LMO with {\bf H}$\| b$. $E_g$(0) values for $\#$1 were obtained from activated-type fits to $\rho_b(T)$ as illustrated in Fig.~\ref{Fig1}a, the $T_{\rm max}$ values were obtained for $\#$2 from the maxima in $\rho_b(T)$ (see Fig.~2c) whilst the $T_{\rm mid}$ values were obtained from the midpoint of the downturns in $\rho_b(T)$. The dashed blue line is a mean-field fit to the $E_g$(0) data. The straight dashed lines are guides to the eye.} \label{Fig2}
\end{figure*}

The activated behavior of $\rho_b(T)$ and the large negative MR for {\bf H}$\| b$ are reminiscent of features seen in the quasi-1D CDW compound (Per)$_2$Au(mnt)$_2$ \cite{Graf04}, where Zeeman splitting of the bands at the Fermi level reduces the pairing interaction and the CDW gap magnitude \cite{Dieterich}. The predicted form of the low-field MR in the CDW state ($T \ll T_{\rm min}$) is \cite{Tiedje}

\begin{eqnarray}
\frac{\Delta \rho_b}{\rho_b} = - \frac{1}{2}\left(\frac{\mu_B H}{k_B T}\right)^2 + {\it 0}\left(\frac{\mu_B H}{k_B
T}\right)^4 \label{eq:one}
\end{eqnarray}
\\
Fig.~1c  shows $\Delta \rho_{b}/\rho_{b}$ of sample $\#$1 for {\bf H}$\| b$ and 0.7 K $\leq T \leq$ 26 K. The development of the negative MR with decreasing temperature is clearly evident from this plot. As indicated by dots in the Figure, the above expression accurately describes the form of the low-field MR at all temperatures $T <$ 10~K, albeit with prefactors that are lower than predicted by a factor of four. More importantly, as shown in Fig.~1d, the data are found to scale with $(\mu_B H/k_B T)^2$ over a decade in $T$. Given the simplicity of the model (single, field-independent gap), this excellent agreement with the model is compelling evidence that gap suppression due to Zeeman splitting completely describes the MR response in LMO for {\bf H}$\| b$.

In a perpendicular field, orbital effects also influence the residual carriers. The magnitude of the MR in LMO for {\bf H}$\| c$ (Fig.~1b) however is too large to be simply a continuation of the conventional orbital MR seen above $T_{\rm min}$. In a quasi-1D system, the Lorentz force induces a sinusoidal real-space modulation of the carrier trajectory along the chains with an amplitude in the orthogonal direction that shrinks as $H$ increases \cite{Gor'kovLebed}. This effective one-dimensionalization can have one of two effects; an increase in the effective nesting of the opposing Fermi sheets \cite{Balseiro}, or a transition to a pure 1D state in which impurities have a much more profound effect on the localization of the (remnant) carriers \cite{Dupuis92}. Both processes are predicted to induce a large positive MR for {\bf I}$\| b$. When combined with the (isotropic) Zeeman effect that causes the gap to collapse at sufficiently high fields, both the orientation and non-monotonic field dependence of $\Delta \rho_{b}/\rho_{b}$ in LMO can be understood on a qualitative level.

This coherent description of the MR response in LMO in terms of CDW suppression appears at odds with the lack of evidence from structural \cite{Santos07}, thermodynamic \cite{Matsuda} or optical studies \cite{Choi04, Degiorgi} of a genuine phase transition in LMO at $T = T_{\rm min}$. From recent thermal expansion studies, dos Santos {\it et al.} have concluded that the non-metallic response in LMO is due to a CDW instability dominated by {\it electronic} interactions, rather than the more conventional phonon-induced (Peierls) instability \cite{Santos07}. As a result, no major structural modification occurs below $T_{\rm min}$.  A strictly 1D Tomonaga-Luttinger liquid with commensurate filling is also susceptible to a transition to a Mott insulating state, and similarly, the opening of a Mott gap might also give rise to an activated form of the resistivity \cite{Giamarchi04} in the absence of a structural transition. However, at present, relatively little is known about how the Mott gap in a 1D system evolves in a magnetic field.

At the lowest temperatures studied, $\rho_b$ of $\#$1 is seen to fall by more than two orders of magnitude for {\bf H}$\| b$. As shown in Fig.~2a, this is sufficient to drive the system metallic, in the sense that $\rho_b(T)$ extrapolates to a finite value as $T \rightarrow$ 0 K. To our knowledge, such field-induced metallization, observed in three different crystals, has never been observed in any other quasi-1D system.

Finally, in sample $\#$2, a striking new feature in the physics of LMO is uncovered beyond the field range where the insulating state is suppressed. Fig.~2b shows $\rho_b(T)$ of sample $\#$2 measured in zero field and an applied field of 45 Tesla ({\bf H}$\| b$). A sharp downturn in the high-field resistivity is observed below a temperature $T_{\rm max} >$ 10 K, located by an arrow in Fig.~2b. These data are re-plotted logarithmically in the inset of Fig.~2b. As shown here and more clearly in the inset of Fig.~2c, the resistivity falls below the noise floor of our measurements, implying a zero-resistive state at the lowest temperatures and highest fields. The evolution of $\rho_b(T)$ between the resistive and zero-resistive state with increasing field strength is illustrated in the main panel of Fig.~2c \cite{secondrun}. The presence of additional maxima and minima at intermediate fields suggests the existence of two competing phases, e.g. density wave formation and superconductivity, as opposed to a more prosaic insulator/metal crossover.

This putative superconducting state is highly dependent on the orientation of the applied field. As shown in Fig.~2d, for example, rotating the field 10$^{\circ}$ towards the $c$-axis is sufficient to restore the resistive state at low $T$. Recall that for {\bf H}$\| a$ and {\bf H}$\| c$, LMO remains insulating in fields up to 30 T (Fig.~1b). This highly directional state contrasts with the case of (TMTSF)$_2$PF$_6$ where pressure induces a superconducting state in zero-field, albeit with anisotropic superconducting parameters \cite{Lee02}.

The resultant phase diagram, derived collectively from measurements on samples $\#$1 and $\#$2, is summarized in Fig.~2e. The gap magnitude $E_g$(0) in $\#$1, extracted from the $T$-dependence of $\rho_b(T)$ at different fields (Fig.~2a), is plotted as blue circles in Fig.~2e. A mean-field-like suppression is observed with a critical field of order 35 Tesla, consistent with the theoretical prediction $\Delta E_g$(0)/$E_g(0) \sim (\mu_B H/2k_B T_{\rm min})^2$ \cite{Tiedje}.  $T_{\rm max}$ and $T_{\rm mid}$, the temperature of the mid-point of the transition, are plotted for each field value as black squares and triangles respectively in Fig.~2e. If $T_{\rm max}$ is indeed associated with the onset of superconductivity, the onset temperature $T_c \sim T_{\rm max}$ is significantly enhanced over the corresponding value in zero-field and intriguingly, appears to {\it increase} with increasing field.  The field at which superconductivity survives is also well above the Pauli paramagnetic limit expected for a 10 K BCS superconductor ($\sim 1.8 k_BT_c/\mu_B$ \cite{Zuo00}), implying either strong spin-orbit scattering or an unconventional order parameter.

Field-induced insulator-superconductor transitions have also been seen in materials, e.g. $\lambda$-(BETS)$_2$FeCl$_4$~\cite{Balicas01}, with a strong internal exchange field that cancels the depairing effect of the externally applied magnetic field - the so-called Jaccarino-Peter compensation effect. However, susceptibility measurements in LMO show no evidence for such strong internal fields ~\cite{Matsuda}. Intriguingly, the recovery of superconductivity and the positive slope in d$T_{\rm max}$/d$H$ are both consistent with theoretical predictions for the high-field behavior of a re-entrant quasi-1D superconductor, including those with triplet pairing~\cite{Lebed86, Dupuis93}. We note that in LMO~\cite{Escribe-Filippini}, the application of pressure leads to a insulator/metal crossover {\it without} significant enhancement of $T_c$, indicating that magnetic fields influence the ground state differently. In the former, pressure presumably enhances the three-dimensionality of the system, whilst in a parallel magnetic field, Zeeman splitting of the Fermi surface(s) suppresses the $2k_f$ density wave without a concomitant change in the dimensionality. In this regard, it would be instructive to learn whether or not the order parameter in zero-field and in the putative high-field superconducting state are in fact the same.

The authors acknowledge J.~F.~Annett, L. Balicas, S.~Carr, D.~Efremov, S.~Kivelson, A.~J.~Schofield and J.~A.~Wilson for stimulating discussions and M.~Greenblatt, J.~P.~A.~Charmant and N.~Fox for collaborative support. This work was supported by the EPSRC (UK), EuroMagNET under EU contract RII3-CT-2004-506239, and a co-operative agreement between the State of Florida and NSF. J.H. and S.Z. acknowledge support from the DOE (DE-FG02-04ER-46139) and SC EPSCoR/Clemson University Cost Share. Research at ORNL is sponsored by the Division of Materials Sciences and Engineering, US DOE.



\begin{thebibliography}{32}
\bibitem{secondrun} The data in Fig. 2c were recorded several months after those shown in Fig. 2b. We note that in the later experiment, the absolute value of $\rho_b(T)$ above 10 K is $\sim 10\%$ higher while the temperature of the maximum in $\rho_b(T)$ (at 45 T) has fallen by $\sim$ 3 K. We attribute this to a slight deterioration in the electrical contacts (causing mixing of the current paths) and possible misorientation of the magnetic field on the second run.
\end{thebibliography}
\end{document}